\documentclass[final,1p,times]{elsarticle} 

\usepackage{graphicx}
\usepackage{amssymb} 
\usepackage{amsthm} 
\usepackage{lineno}

\journal{Nuclear Physics A} 

\begin{document}

\begin{frontmatter} 

\title{PHENIX Highlights}

\author{Takao Sakaguchi (for the PHENIX\fnref{col1} Collaboration)}
\fntext[col1] {A list of members of the PHENIX Collaboration and acknowledgements can be found at the end of this issue.}
\address{Brookhaven National Laboratory, Physics Department, Upton, NY 11973, USA}


\begin{abstract} 
PHENIX reports on electromagnetic and hadronic observables in large
data sets of p+p, d+Au and Au+Au collisions at various
cms energies. Initial state effects in cold nuclear matter are 
quantified by centrality dependent $\pi^0$, $\eta$, reconstructed
jets and $\psi\prime$ measurements. 
Using the first run of the new EBIS ion source at RHIC, we report first
results for particle flow ($v_1$ and $v_2$) and quarkonium production
in U+U and Cu+Au collisions. 
Hot matter
created in Au+Au is characterized using event-plane
dependent HBT and dielectrons. 
Parton-medium interactions are investigated using high $p_T$ single
hadrons, $\gamma$-hadron correlations and 
heavy flavor decay electrons identified with the newly installed VTX detector.
\end{abstract} 

\end{frontmatter} 


\section{Introduction}
RHIC experiments have demonstrated that quark-gluon plasma (QGP) is created
in Au+Au collisions.
Subsequent experimental goals 
focus on characterizing the QGP, and exploring the phase transition
and a possible critical point
by varying the colliding nuclei and collision energy.
We present the latest results from the PHENIX experiment at RHIC.
The results are categorized according to time within the collision: 
initial state effects, hot matter dynamics, and parton-medium interactions.

\section{The baseline for QGP properties}
Initial state effects on the primordial hard scatterings which produce
hard probes of the hot dense medium reflect parton distribution
functions in nucleons and nuclei, as well as possible energy loss or 
bound state dissociation in cold matter. p+p collisions provide the baseline, while d+Au collisions reflect additional cold nuclear matter (CNM) effects arising from the presence of a nucleus in the collision. Direct photon spectra
in d+Au collisions show little or no modification 
compared to expectations from p+p collisions, though an
isospin effect is consistent with the data~\cite{bib0,bib1}.
On the other hand, hadronic probes display centrality-dependent CNM effects.

\begin{figure}[ht]
\begin{center}
\includegraphics[width=0.8\textwidth]{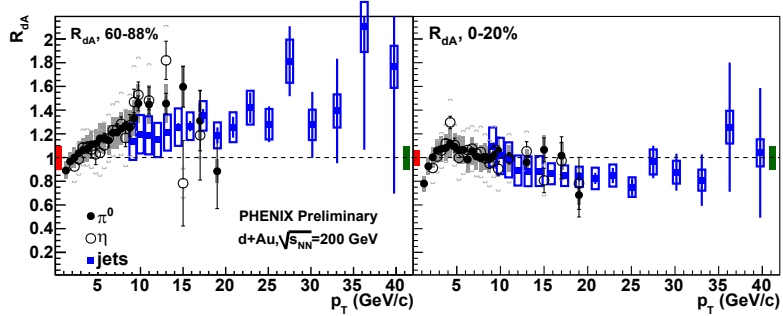}
\caption{Nuclear modification factors ($R_{dA}$) for $\pi^0$, $\eta$ and fully-reconstructed jets in d+Au collisions.  \label{RAA_pi0_eta_jets} }
\end{center}
\end{figure}

In Fig.~\ref{RAA_pi0_eta_jets}, the nuclear modification
factors ($R_{dA}$) for $\pi^0$, $\eta$ and
fully-reconstructed jets in d+Au collisions at $\sqrt{s_{NN}}$=200\,GeV
are shown~\cite{bib2}.
These three probes result from different analysis techniques, but
show remarkable consistency. While a slight suppression is
found in the most central collisions, there is significant enhancement
at high $p_T$ in the
peripheral collisions; such a feature is not predicted by available 
models. We note that centrality is determined in the data
by the particle multiplicity in the beam-beam counter (BBC)
at 3.1$<|\eta|<$3.9, while the models use impact parameters.
Therefore, the centrality classes between data and models may differ somewhat,
particularly for events with high $p_T$ hadrons or jets ($p_T>$10\,GeV).

\begin{figure}[ht]
\begin{minipage}{62mm}
\begin{center}
\includegraphics[width=1.0\textwidth]{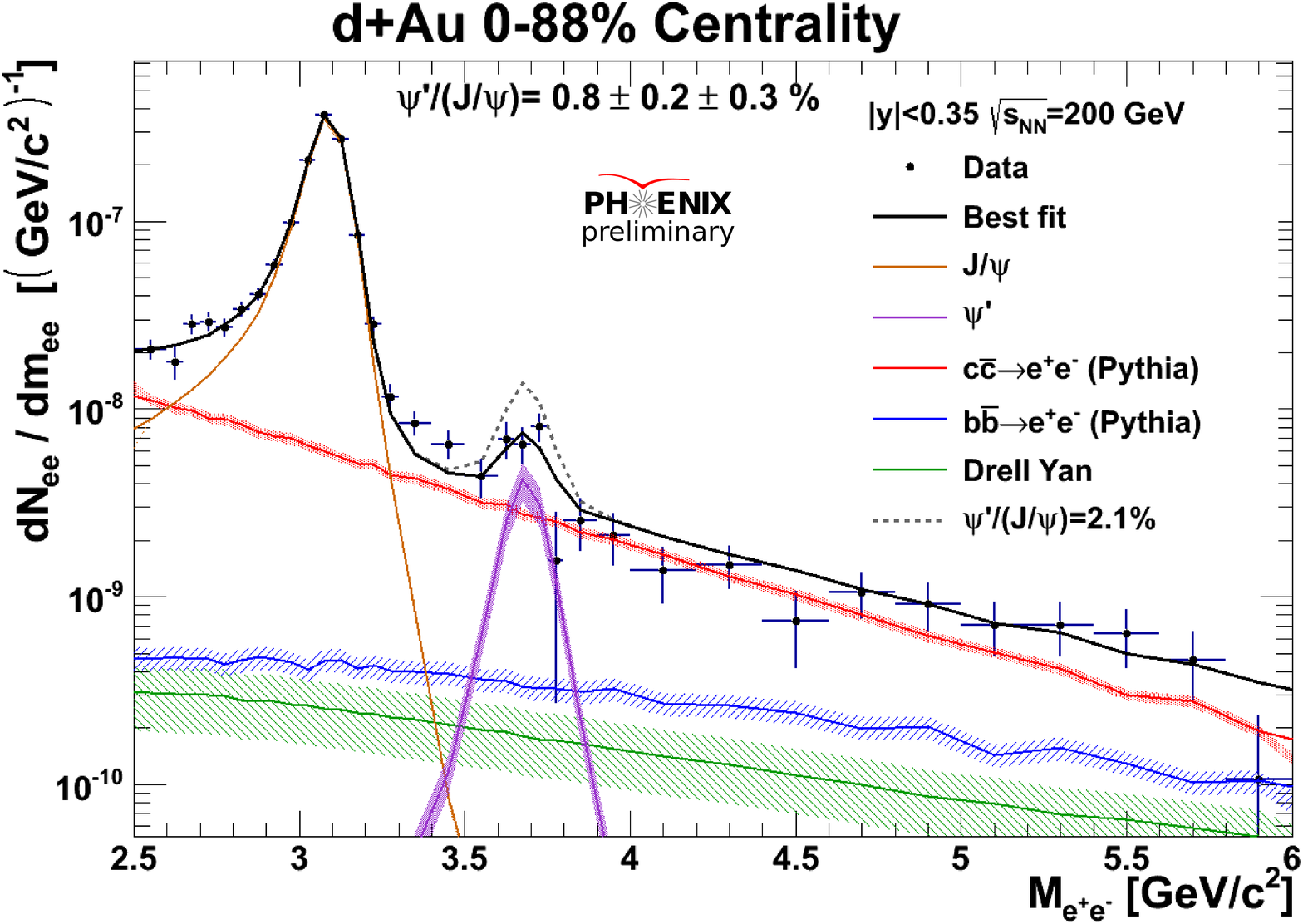}
\caption{$\psi\prime$ peak seen in dielectron mass spectra in minimum bias d+Au collisions. \label{PsiPrime_Mass} }
\end{center}
\end{minipage}
\hspace{6mm}
\begin{minipage}{62mm}
\begin{center}
\includegraphics[width=1.0\textwidth]{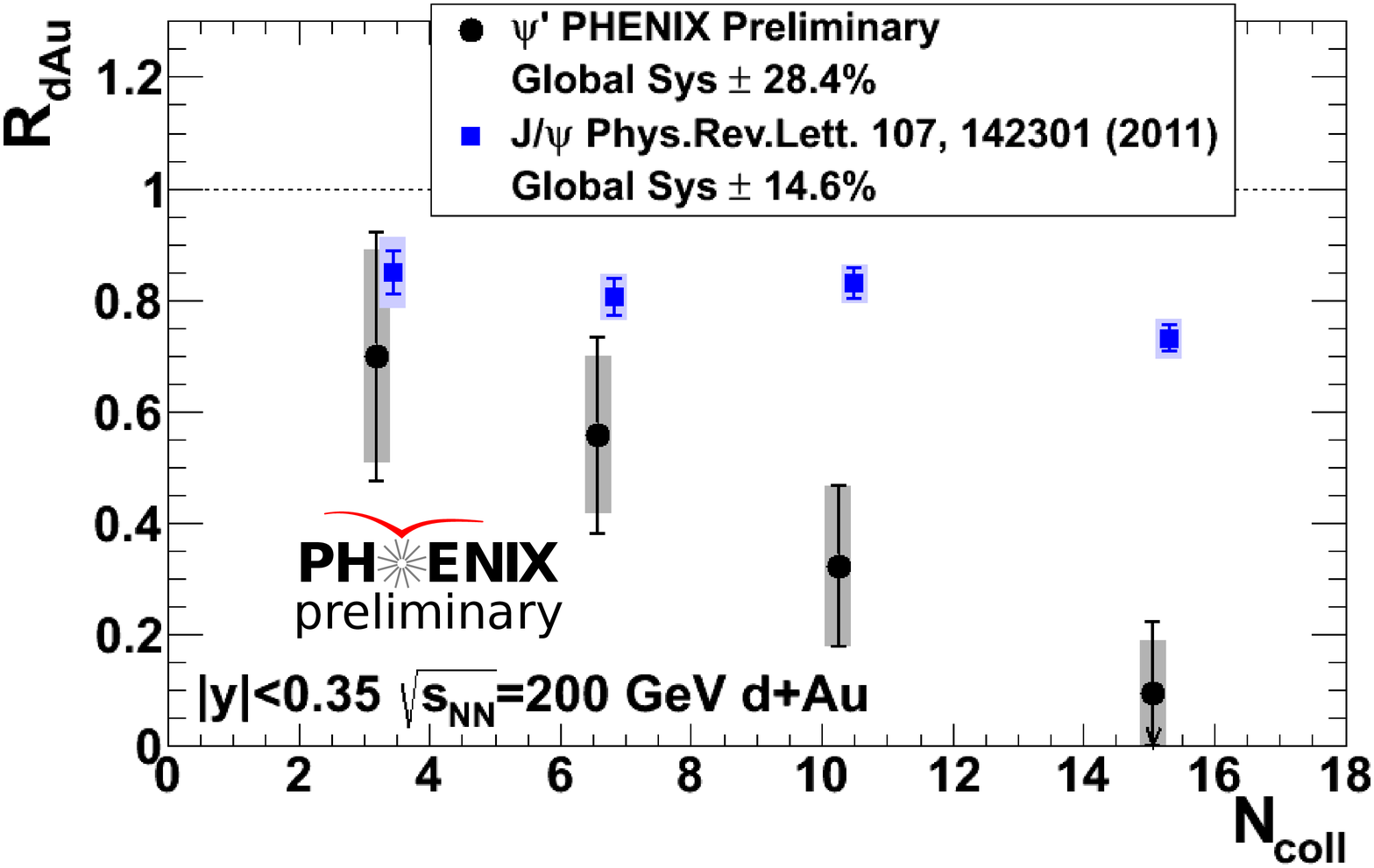}
\caption{$R_{dA}$ of $\psi\prime$ and $J/\psi$ in d+Au collisions as a function of centrality. $\psi\prime$ is more suppressed than $J/\psi$.  \label{PsiToJPsi} }
\end{center}
\end{minipage}
\end{figure}

Quarkonia offer key probes of initial state effects. The
high statistics 2008 data enable measurement of $\psi\prime$
production as a function of collision centrality in d+Au at $\sqrt{s_{NN}}$=200\,GeV. The dielectron mass spectrum in the
$J/\psi$ and $\psi\prime$ region is shown in
Fig.~\ref{PsiPrime_Mass}~\cite{bib2}.
The ratio of $\psi\prime/(J/\psi)$ is $\sim0.8$\,\%, smaller than in p+p (2.1\,\%). The centrality dependence of the
ratio is shown in Fig.~\ref{PsiToJPsi}. The yield of $J/\psi$
is somewhat suppressed in d+Au collisions compared to p+p, but 
$\psi\prime$ is more heavily suppressed. The relative suppression 
at FNAL fixed target energy can be reasonably well explained by considering
the time the precursor $c \overline{c}$ state spends in the nucleus. 
However, this explanation completely fails at RHIC energy.

\section{New collision systems for QGP characterization}
In 2012, first data were taken with the EBIS ion source, providing
new ion species at RHIC.

\subsection{U+U collisions}
\begin{figure}[ht]
\begin{minipage}{68mm}
\begin{center}
\includegraphics[width=1.0\textwidth]{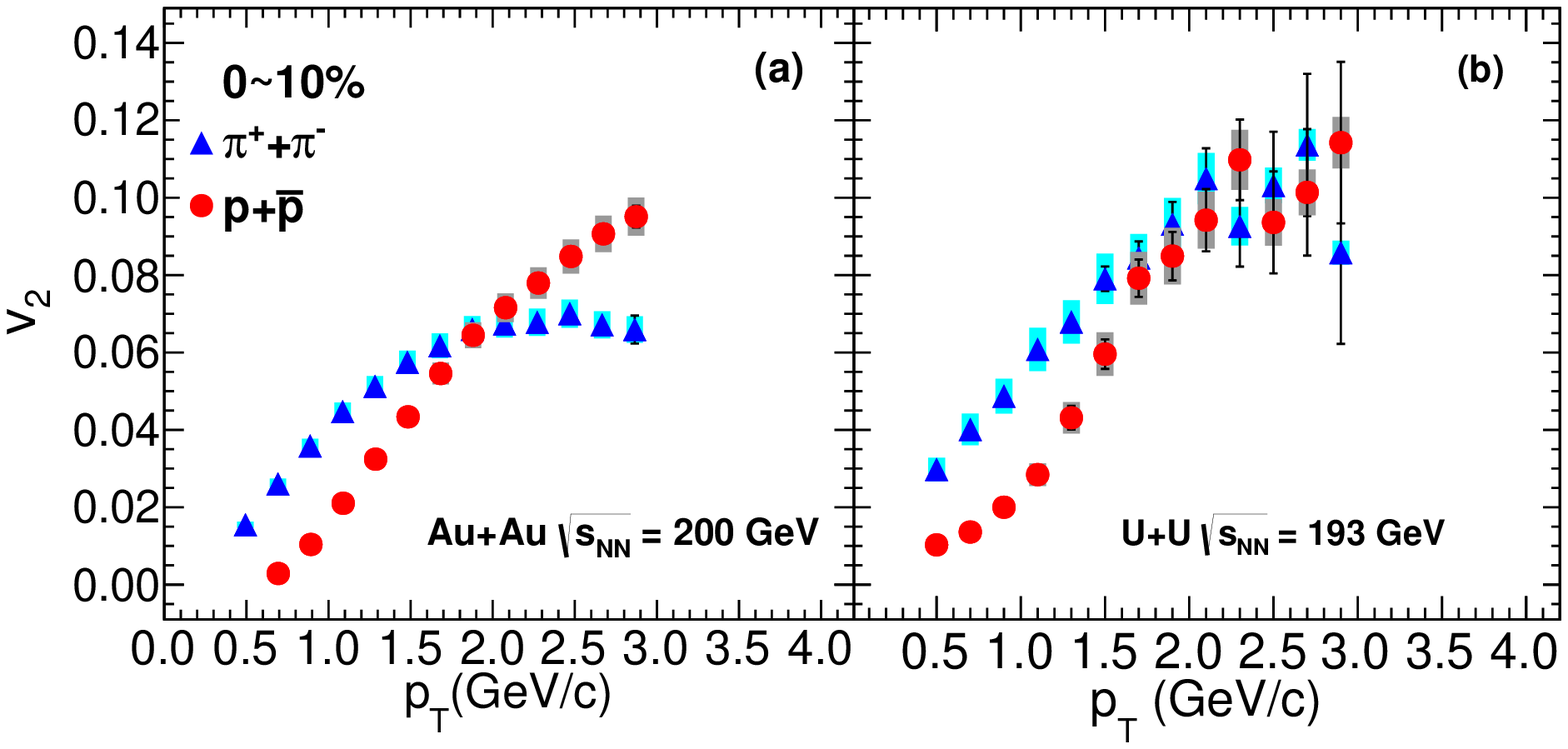}
\caption{Elliptic flow of charged pions and protons in (a) Au+Au and (b) U+U collisions at 0-10\,\% centrality.  \label{UU_v2} }
\end{center}
\end{minipage}
\hspace{3mm}
\begin{minipage}{60mm}
\begin{center}
\includegraphics[width=1.0\textwidth]{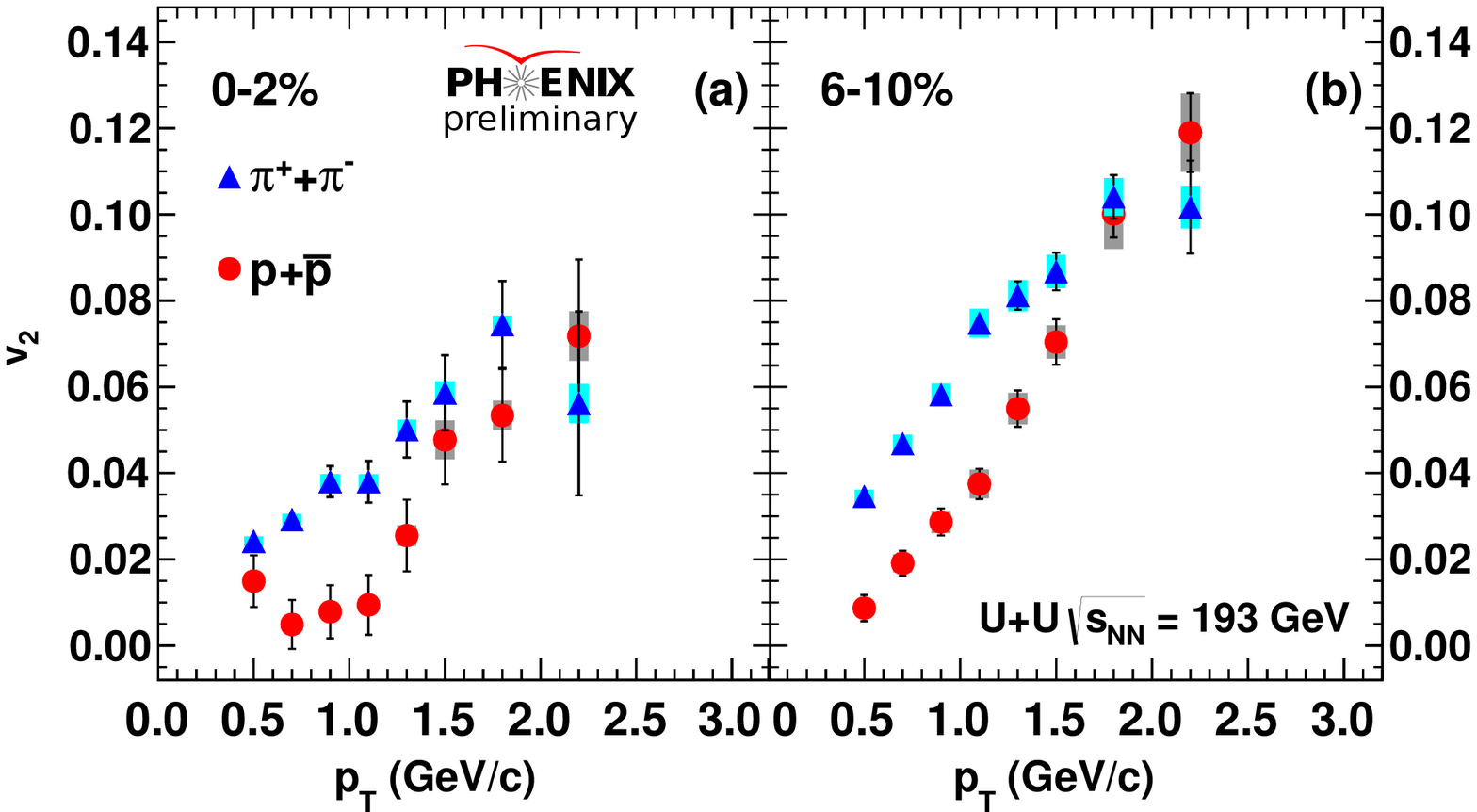}
\caption{Elliptic flow of charged pions and protons in (a) 0-2\,\% and (b) 6-10\,\% centrality in U+U collisions.  \label{UU_v2_fine} }
\end{center}
\end{minipage}
\end{figure}

Uranium is a highly deformed nucleus. Therefore, selecting
collisions where two uranium nuclei are aligning with their
long axes head-on (tip-tip)
should access higher energy density at RHIC energy.
PHENIX collected data in U+U collisions at $\sqrt{s_{NN}}$=193\,GeV.
First results show that the Bjorken energy
density for 0-2\,\% centrality (tip-tip enriched events) is $\sim$30\,\%
higher than the most central Au+Au collisions at 200GeV~\cite{bib3}.
The elliptic flow of identified charged pions and protons 
in U+U collisions are compared to that in Au+Au at 0-10\,\% 
centrality in Fig.~\ref{UU_v2}~\cite{bib4}.
For pions in U+U, $v_2$ is flatter with $p_T$ for
$p_T<$1\,GeV, 
reminiscent of observations in Pb+Pb at LHC~\cite{bib5}. This can be 
interpreted as a consequence
of very strong radial flow for higher energy density matter. 
The finer centrality classes shown in Fig.~\ref{UU_v2_fine} show that
this flattening is most pronounced in 0-2\,\%, where tip-tip events
represent a larger fraction of the total, and is nearly nonexistent 
for 6-10\,\% centrality.
It is intriguing to see this phenomenon at RHIC, as the energy density
reached at LHC is $\sim$3 times higher than at RHIC.

\subsection{Cu+Au collisions}
\begin{figure}[ht]
\begin{minipage}{62mm}
\begin{center}
\includegraphics[width=0.75\textwidth]{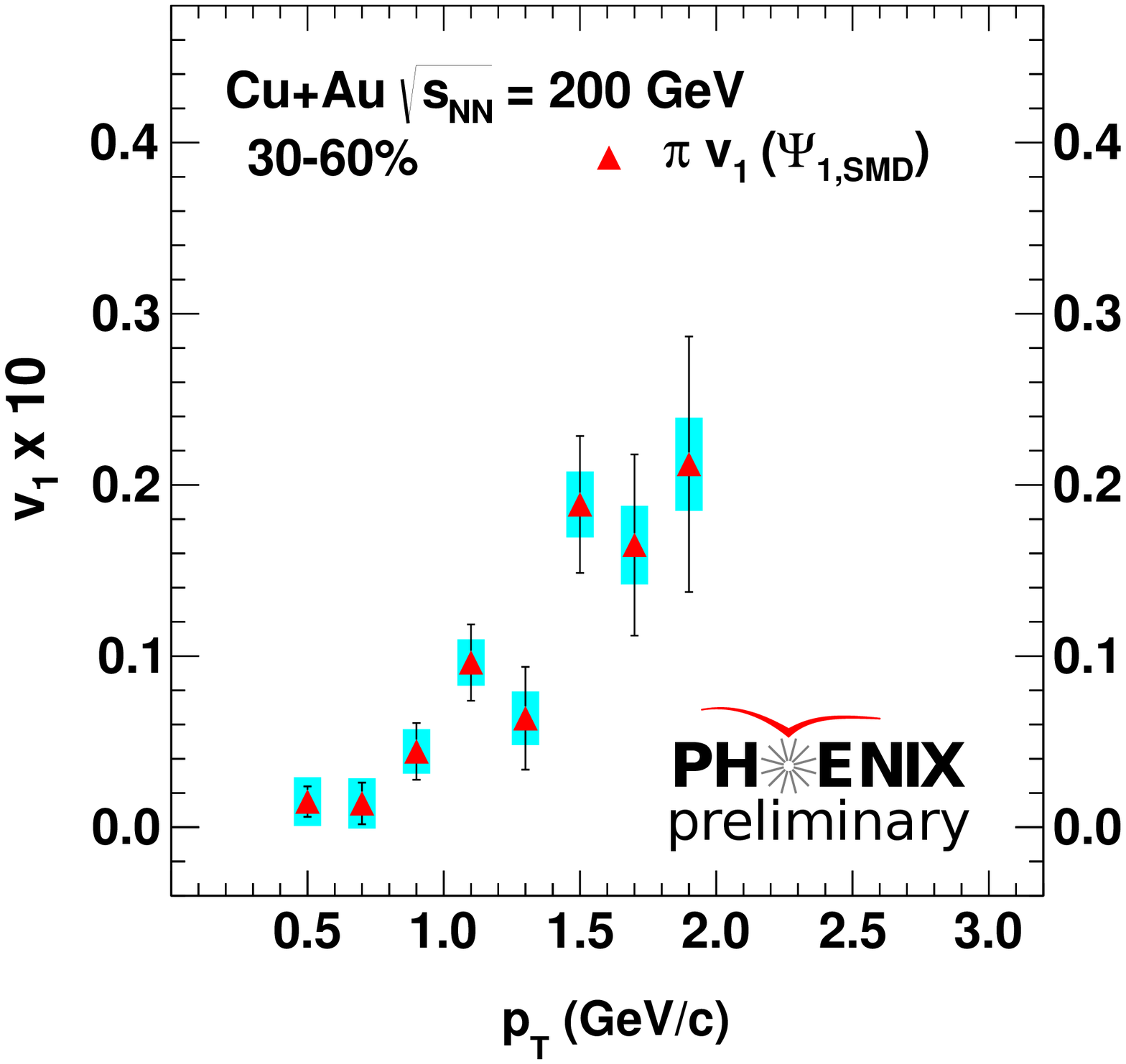}
\caption{$v_1$ of charged pions in Cu+Au collisions.  \label{v1_CuAu} }
\end{center}
\end{minipage}
\hspace{5mm}
\begin{minipage}{62mm}
\begin{center}
\includegraphics[width=0.8\textwidth]{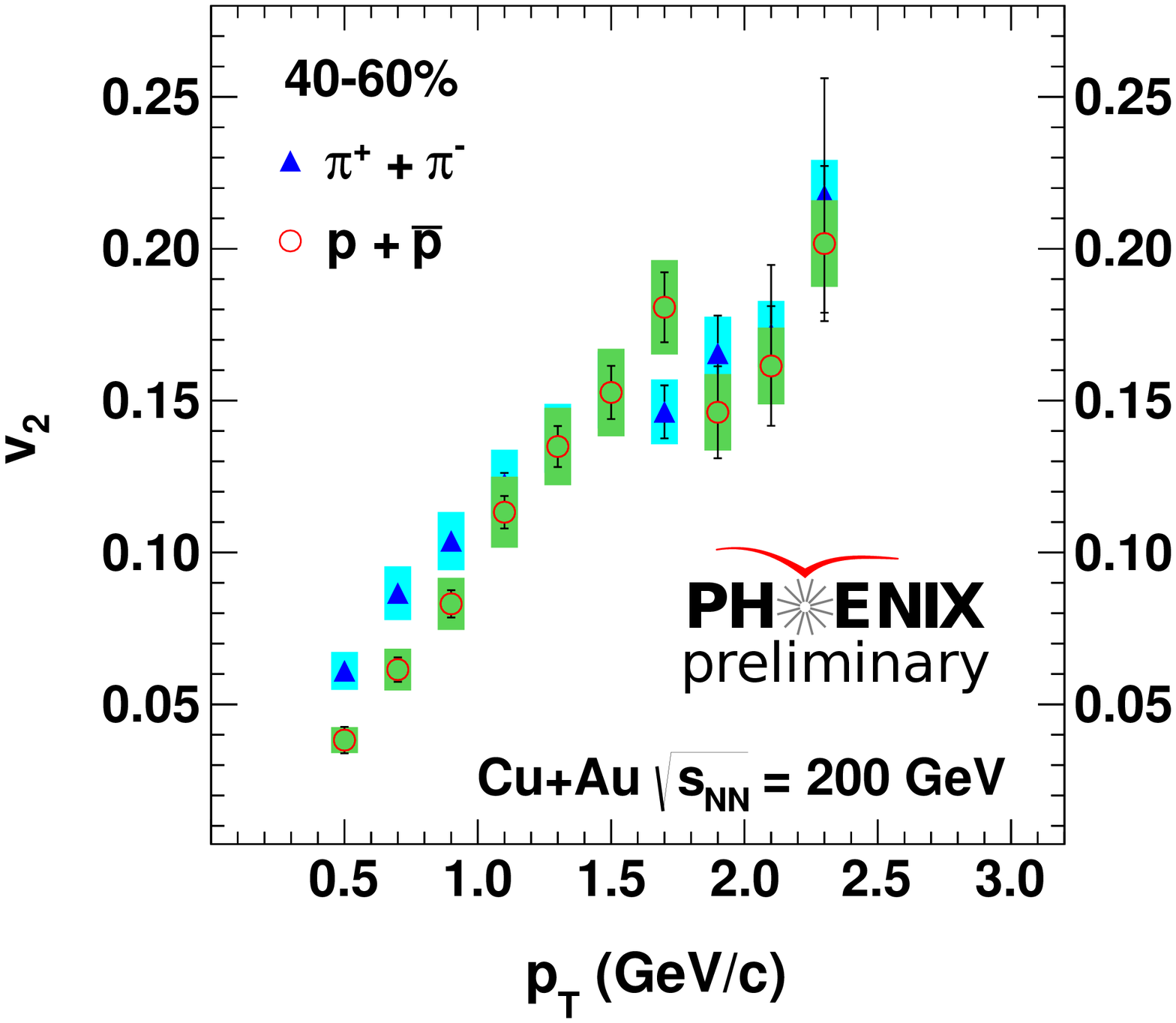}
\caption{$v_2$ of charged pions and protons in Cu+Au collisions.  \label{v2_CuAu} }
\end{center}
\end{minipage}
\end{figure}
In Cu+Au
collisions, one can select events in which the Cu is fully contained 
within the Au
nucleus. Furthermore, the pressure gradient developed 
should be asymmetric, resulting in an intrinsic
triangularity in the collective motion. This triangularity generates
a $v_3$ that is geometric, rather than fluctuation driven. PHENIX 
has already analyzed $\sim$20\,\% of the 2012 Cu+Au data. 
The shower-max detector in the zero-degree calorimeters is used to determine
the event plane (reaction plane $\Psi_1$), allowing a measurement of $v_1$.
We defined the Au-spectator going side as positive $\Psi_1$. 

Fig.~\ref{v1_CuAu} and \ref{v2_CuAu} show $v_1$ and $v_2$ for 
identified pions and protons.
A sizable positive $v_1$ is observed;
the sign and magnitude of $v_1$ differ from predictions of the
the AMPT model (HIJING+parton cascade)~\cite{AMPTref}, although
AMPT describes
symmetric collision systems very well.

\begin{figure}[ht]
\begin{center}
\includegraphics[width=0.70\textwidth, trim=5 5 5 5, clip]{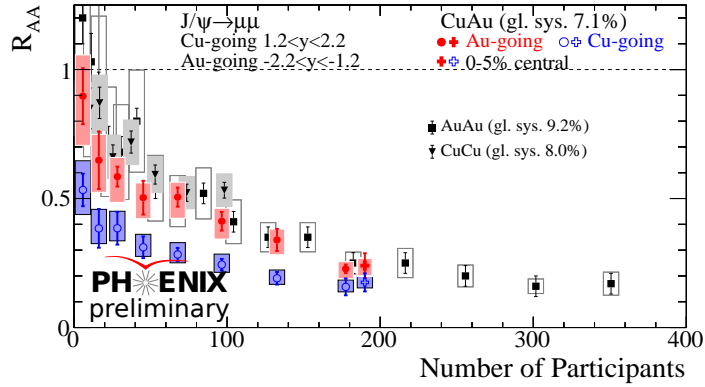}
\end{center}
\caption{$R_{AA}$ of $J/\psi$ in Cu+Au, Cu+Cu and Au+Au collisions.}
\label{Jpsi_CuAu}
\end{figure}

Utilizing the muon arms at 1.2$<|\eta|<$2.2, PHENIX measured
$J/\psi$ spectra in Cu+Au collisions via the di-muon decay channel. This
allowed determination of $J/\psi$ $R_{AA}$ in Cu+Au, and comparison to the
the $J/\psi$ suppression observed in Cu+Cu and Au+Au collisions.
The results and comparison are shown in
Fig.~\ref{Jpsi_CuAu}~\cite{bib6}.
The suppression of the $J/\psi$ yield as a function of $N_{part}$ on the 
Au-going
side fits well with the $N_{part}$ dependence in Au+Au and Cu+Cu collisions.
However, on the
Cu-going side, the suppression is significantly stronger. The difference of
the nuclear PDFs in Cu and Au nuclei partly explains this 
trend, but additional final state effects may need to be
considered in order to understand the magnitude on the Cu-going side.

\section{Characterizing hot dense matter by new probes}
Two boson correlations as pioneered by Hanbury-Brown and Twiss (colloquially known as HBT correlations in our field) provide information on the 
space-time evolution of the
particle emission source. PHENIX has measured the angular dependence of
HBT radii with respect to the (second order) event plane ($R^{2}_{s} (\Delta\phi) = R^{2}_{s,0}+2\times\Sigma R^{2}_{s,n} cos(n\Delta\phi)$). This analysis shows
that the eccentricity, defined as $\epsilon_{HBT}=2R^{2}_{s,2}/R^{2}_{s,0}$,
obtained from the Kaon HBT correlations is consistent with the initial
eccentricities. This implies early freezeout of kaons from the expanding
hadronic phase late in the collision.

\begin{figure}[ht]
\begin{center}
\includegraphics[width=0.8\textwidth]{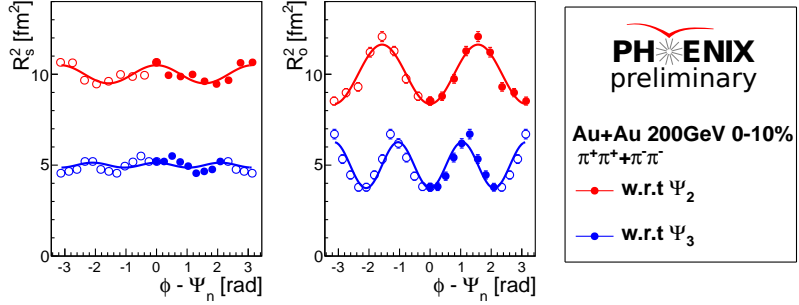}
\end{center}
\caption{Angular dependence of HBT radii ($R_{s}$ and $R_{o}$) with respect to 2nd ($\Psi_2$) and 3rd ($\Psi_3$) order event plane in 0-10\,\% Au+Au collsion at $\sqrt{s_{NN}}$=200\,GeV.}
\label{HBT_2nd_3rd}
\end{figure}

PHENIX studied the HBT radii with respect to the 3rd order
event plane for the first time. Fig.~\ref{HBT_2nd_3rd} shows $R_{s}$ and
$R_{o}$ for charged pions as a function of the emission angle with respect
to the 2nd and 3rd order event plane in 0-10\,\% in Au+Au collisions~\cite{bib7}.
The average radii are 
10 and 5 for $\Psi_2$ and $\Psi_3$, respectively.
While $R_s$ shows only a weak dependence on orientation with respect to the
event plane, $R_o$ indicates a clear
temporal variation as a function of angle for both the $\Psi_2$ and $\Psi_3$
event planes. Moreover, the magnitude of the oscillation for $\Psi_3$ is
almost the same as $\Psi_2$. Since $R_{o}$ includes the emission duration
in addition to the spatial homogeneity length in the transverse direction,
this measurement reflects the space-time evolution of
ellipticity and triangularity of the collision system.

Turning to thermal radiation, we report on a unique and penetrating probe of
thermodynamic quantities. Measuring thermal radiation gives access, for example, to the temperature of the hot dense medium
produced in the collision. Using the Hadron Blind Detector (HBD)
installed in 2010, PHENIX performed a new measurement of 
thermal di-electron radiation. In all bins from peripheral to semi-central, the new result is consistent with that from the 2004 data. The most central bin is still being analyzed.
For further details, see \cite{bib8}.

\section{Detailed study of energy loss of partons}
\subsection{$\gamma$-h correlation}
\begin{figure}[ht]
\begin{center}
\includegraphics[width=0.75\textwidth, trim=0 5 5 5, clip]{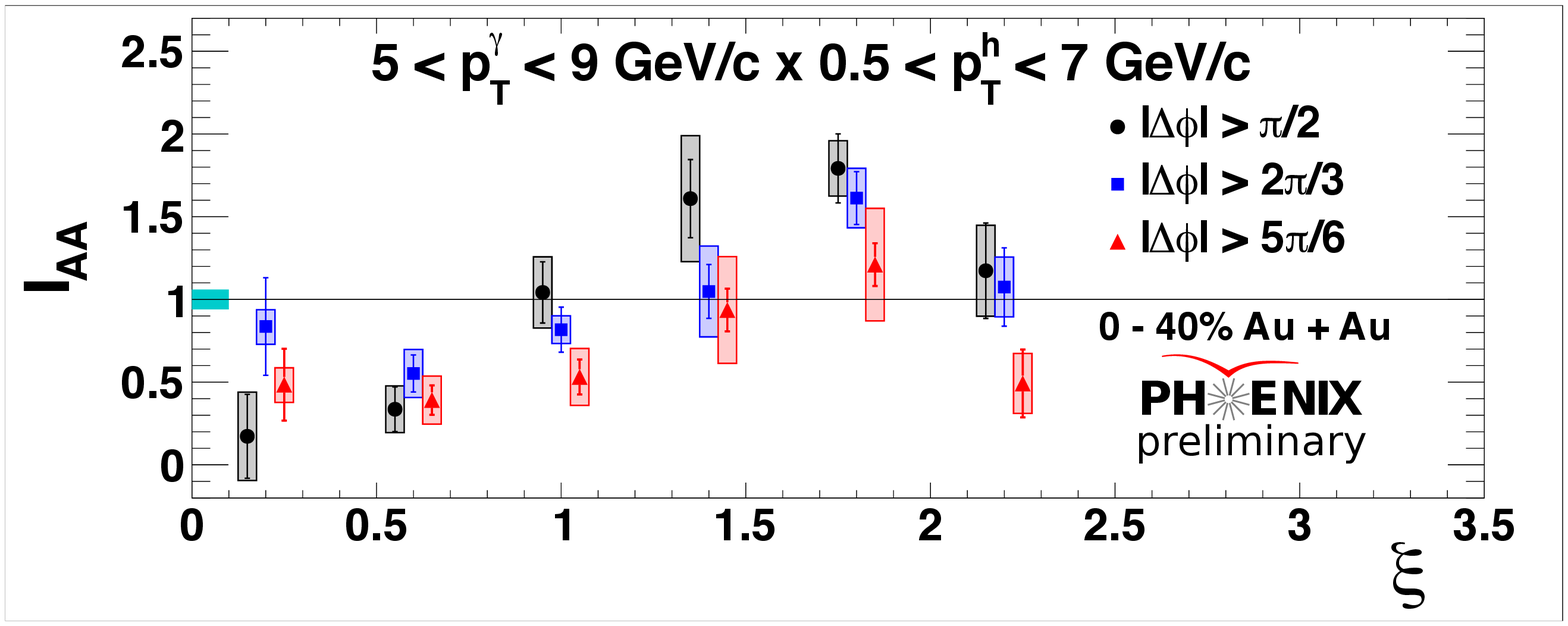}
\end{center}
\caption{Ratio of per-trigger away-side hadron yield in 200\,GeV Au+Au and p+p $\gamma$-hadron measurements as a function of $\xi$. The legend identifies the three different integrating angular ranges.}
\label{Gamma_h}
\end{figure}
$\gamma$-hadron correlations are often considered a "golden channel" to
evaluate the energy loss of partons in the medium. Since direct photons
reflect the momentum of the original scattered partons, they tag the initial
momentum of the scattered parton that traverses the medium. Consequently
one can deduce the energy loss via $\delta p_T = p_T(\gamma)-p_T(hadrons)$.
PHENIX has measured the hadron yield opposing direct photons as a function of
$\xi$($\equiv-ln(z_T)$) for the 2007 and 2010 Au+Au data sets together.
The results are shown in Fig.~\ref{Gamma_h}~\cite{bib9}.
As the angular range in which the away-side hadrons are integrated is
increased from
$|\delta \phi|>5\pi/6$ to $|\delta \phi|>\pi/2$, the per-trigger
away-side yield increases compared to p+p collisions.
The increase is in the larger $\xi$ or small $z_T$ region, implying that the 
softer away-side jet fragments are distributed over a wider angle than
fragments carrying a large fraction of the jet energy. This is expected
from medium-enhanced splitting, and is qualitatively consistent
with the angular broadening observed in hadron-hadron
correlations at RHIC and the jet shape broadening
observed at LHC.

\subsection{High $p_T$ $\pi^0$ to determine fractional momentum shift}
\begin{figure}[ht]
\begin{center}
\includegraphics[width=0.72\textwidth]{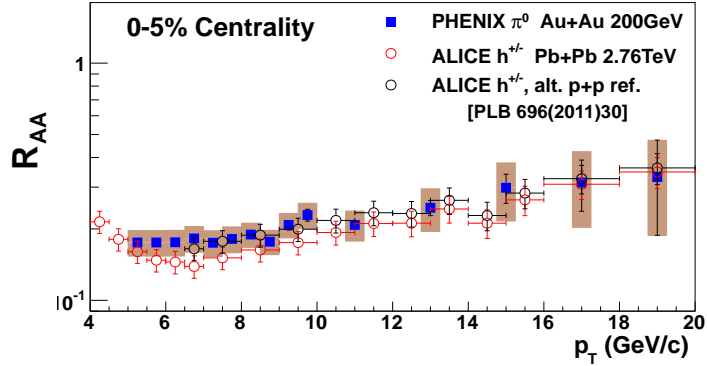}
\caption{Comparison of $R_{AA}$ of $\pi^0$ from PHENIX at RHIC and charged hadrons from ALICE experiments at LHC. Two data sets from ALICE experiments correspond to two different p+p references. \label{ALICE_PHENIX_Comp} }
\end{center}
\end{figure}

\begin{figure}[ht]
\begin{minipage}{58mm}
\begin{center}
\includegraphics[width=0.9\textwidth]{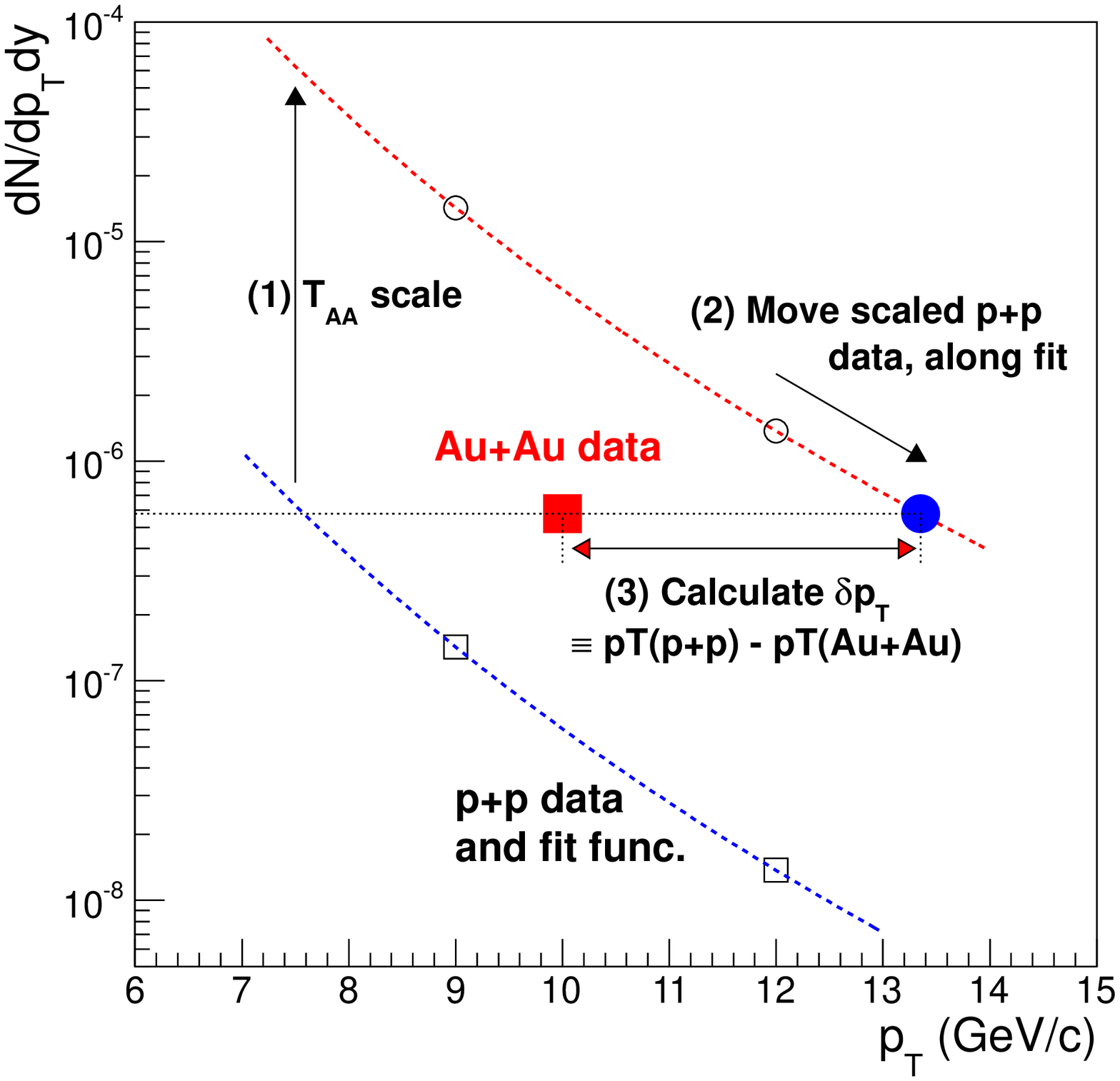}
\caption{Demonstration of calculating the fractional momentum loss ($S_{loss}$).  \label{Sloss_Method} }
\end{center}
\end{minipage}
\hspace{5mm}
\begin{minipage}{67mm}
\begin{center}
\includegraphics[width=1.0\textwidth]{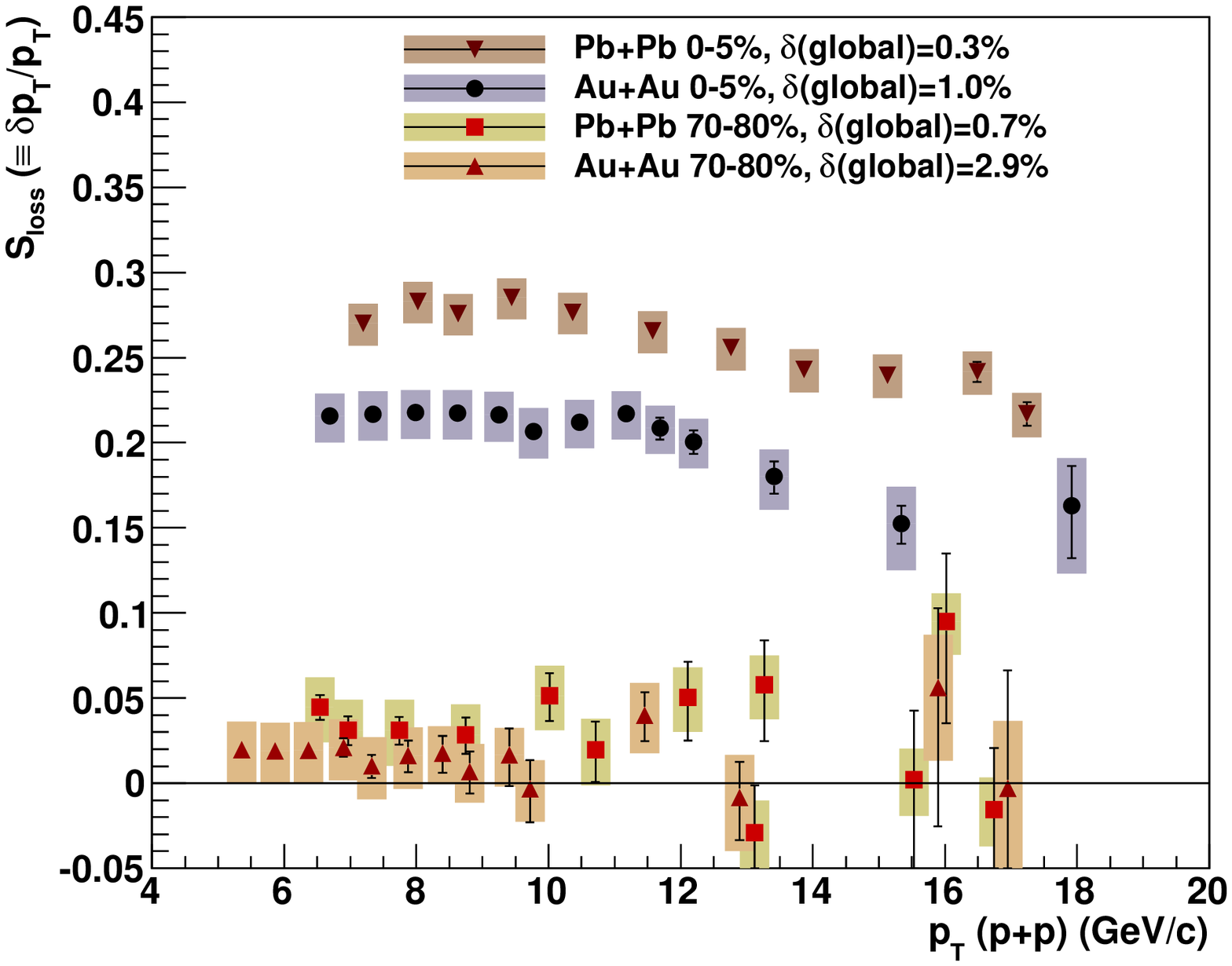}
\caption{$S_{loss}$ for PHENIX $\pi^0$ and ALICE charged hadrons. \label{Sloss_ALICE_PHENIX} }
\end{center}
\end{minipage}
\end{figure}

Using the large 2007 Au+Au data set at $\sqrt{s_{NN}}$ = 200 GeV and a
new algorithm to correct for shower merging, 
PHENIX extended the $p_T$ range
of $\pi^0$ spectra~\cite{bib10}.
Fig.~\ref{ALICE_PHENIX_Comp} compares $\pi^0$ $R_{AA}$ in 
200\,GeV Au+Au collisions from RHIC and charged hadron $R_{AA}$ measured in
2.76\,TeV Pb+Pb at the LHC (ALICE experiment)~\cite{bib11}.
For both centralities and over the entire $p_T$ range shown, the two data
sets are rather similar.
However, this does not necessarily mean that the energy loss of the partons is
the same, as the spectral shapes differ. PHENIX determines the average
fractional momentum shift 
($S_{loss}$) of high $p_T$ $\pi^0$ to estimate the average fractional
energy loss of the initial parton.
$S_{loss}$ is defined as $\delta p_{T}/p_{T}$, where $\delta p_{T}$ is
the difference of the momentum in p+p collisions 
($p_{\rm T,pp}$) and
Au+Au ($p_{\rm T,AuAu}$); $p_T$ in the denominator
is $p_{\rm T,pp}$.
The calculation method is schematically depicted in
Fig.~\ref{Sloss_Method}.
First, the $\pi^0$ cross-section in p+p ($f(p_{\rm T})$) is scaled by
$T_{AA}$ corresponding to the centrality selection of the Au+Au data
($g(p_{\rm T})$). Next, the scaled p+p cross-section
($T_{\rm AA}f(p_{\rm T})$) is fit with
a power-law function ($h(p_{\rm T})$). Third, a scaled p+p point closest in
yield ($p_{\rm T, pp}'$) to the Au+Au point of interest ($p_{\rm T, AuAu}$)
is shifted using the fit function as
\begin{eqnarray}
T_{\rm AA}f(p_{\rm T,pp}) = h(p_{\rm T,pp})/h(p_{\rm T,pp}')
 \times T_{\rm AA}f(p_{\rm T,pp}'),
\end{eqnarray}
where $p_{\rm T, pp}$ is chosen to satisfy the relation
\begin{eqnarray}
T_{\rm AA}f(p_{\rm T,pp}) = g(p_{\rm T,AuAu}).
\end{eqnarray}
The $\delta p_T$ is calculated as 
$p_{\rm T,pp} - p_{\rm T,AuAu}$. For obtaining $S_{loss}$, the
$\delta p_{\rm T}$ is divided by the $p_{\rm T,pp}$.
We show the comparison of $S_{loss}$ for PHENIX $\pi^0$'s in Au+Au and
ALICE charged hadrons in Pb+Pb in Fig.~\ref{Sloss_ALICE_PHENIX}. We took
the corresponding Pb+Pb and p+p data for black points in
Fig.~\ref{ALICE_PHENIX_Comp} for calculating the $S_{loss}$ for ALICE
charged hadrons.
The $S_{loss}$ for the ALICE charged hadrons is found to be $\sim$25\,\%
higher than that for the PHENIX $\pi^0$'s.
PHENIX has also measured $S_{loss}$ in Au+Au at 39 and
62.4\,GeV cms energy. The change in $S_{loss}$ reaches approximately
a factor of 4 from 200\,GeV to 39\,GeV~\cite{bib12,bib13}.

\section{Heavy flavor electrons}
\begin{figure}[htbp]
\begin{minipage}{66mm}
\begin{center}
\includegraphics[width=1.0\textwidth]{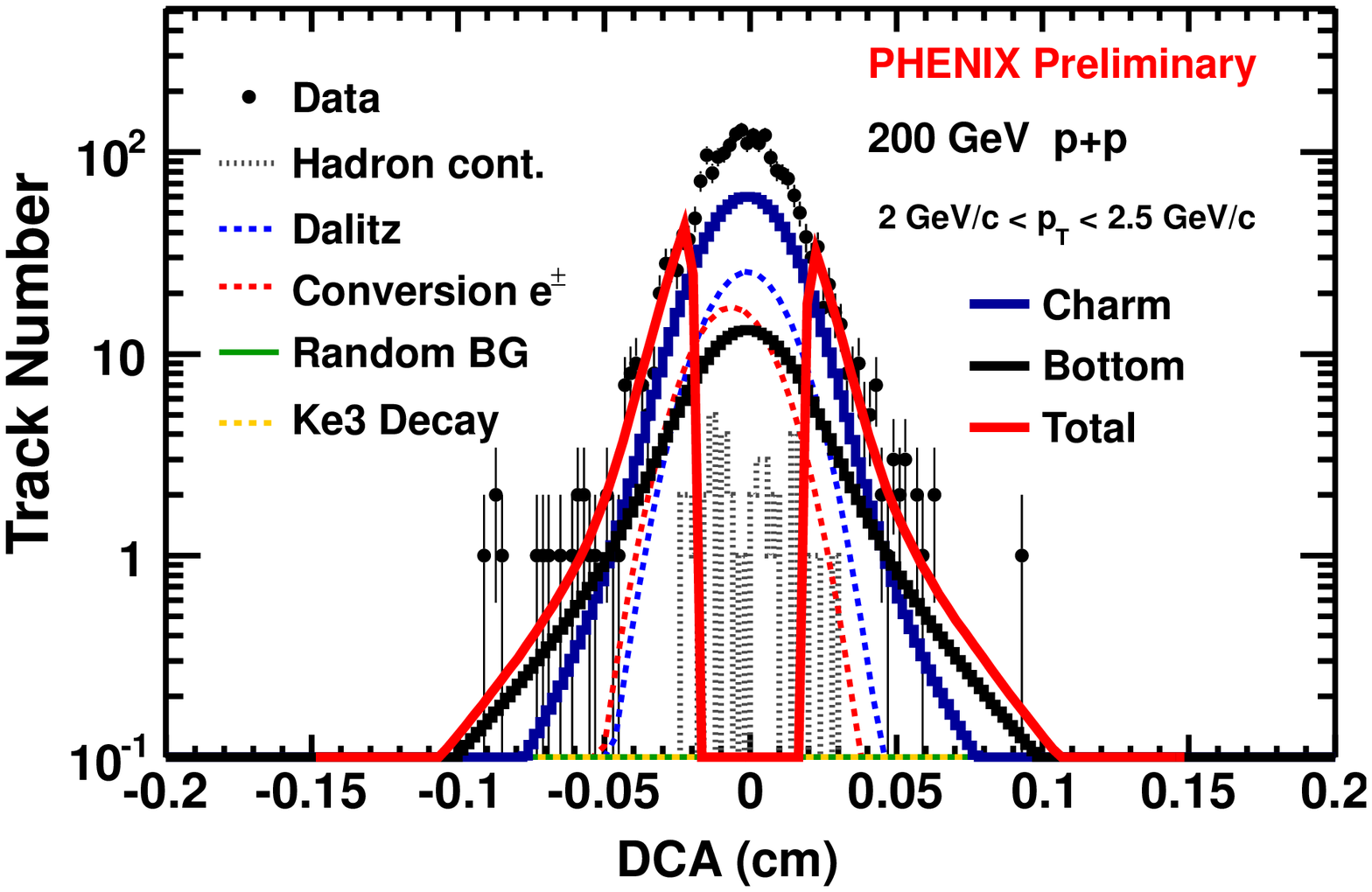}
\end{center}
\vspace{-3mm}
\caption{DCA distribution in p+p collisions for 2$<p^e_T<$2.5\,GeV/$c$, and its deconvolution to contributions from parent hadrons.  \label{VTX_DCA} }
\end{minipage}
\hspace{5mm}
\begin{minipage}{58mm}
\begin{center}
\includegraphics[width=1.0\textwidth]{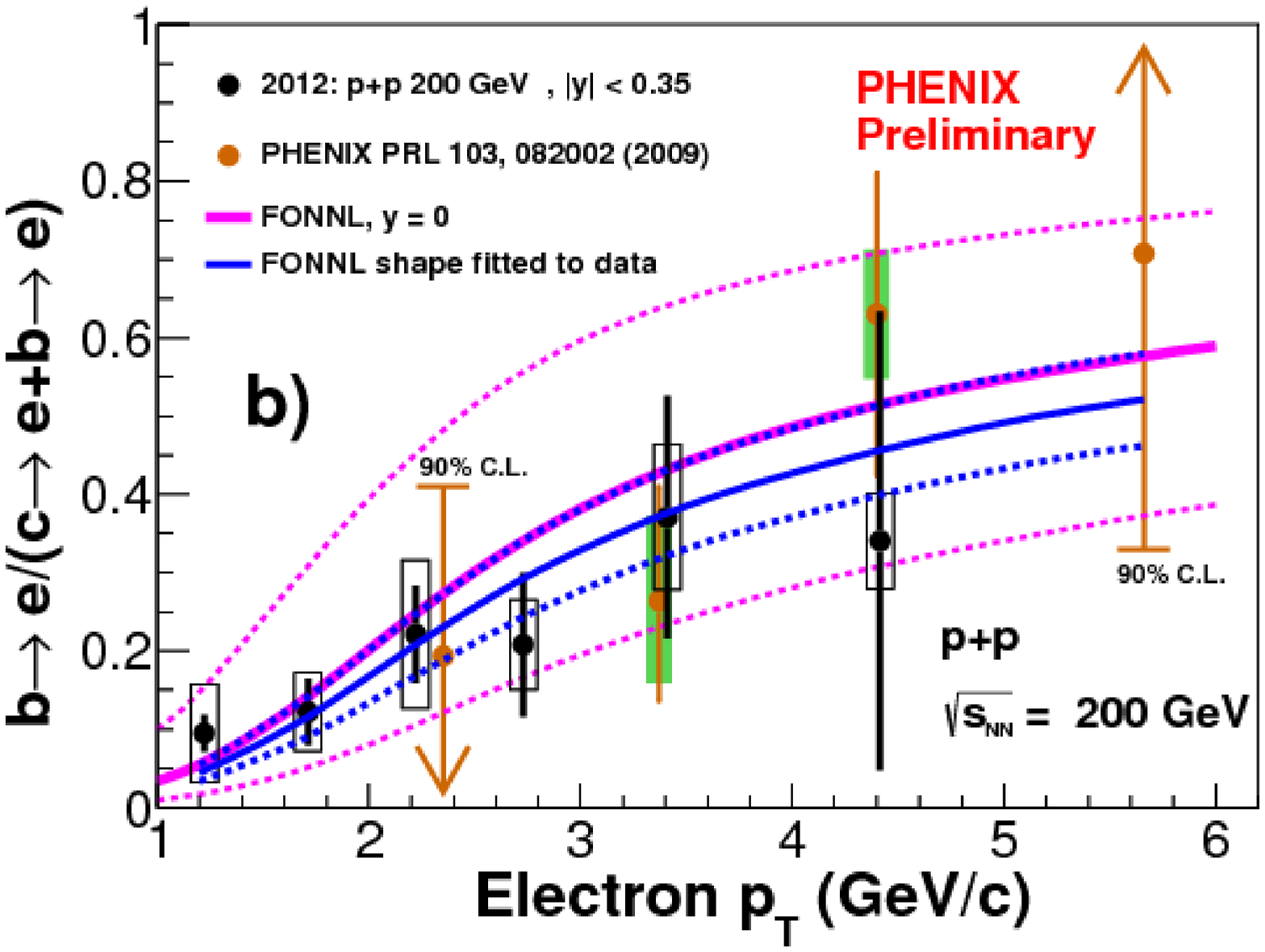}
\end{center}
\vspace{-3mm}
\caption{Ratio of electron yields from bottom quarks to bottom and charm quarks in p+p collisions.  \label{VTX_pp} }
\end{minipage}
\end{figure}

The suppression of non-photonic electron yields 
was found to be nearly as large as that of $\pi^0$'s.
The question has been whether the suppression arises from the
suppression of charm and/or bottom quarks. Theoretical predictions
about this differ greatly. Consequently, PHENIX constructed a micro
vertex detector to separate electrons from charm and bottom decays.
This detector, the VTX, consists of 2 layers each of strip-pixel and pixel silicon detectors.
The measurement of the distance of closest approach (DCA) of electrons
to the collision vertex allows separation according to the parent
hadrons. Fig.~\ref{VTX_DCA} shows a DCA
distribution measured in the VTX detector in p+p collisions.
The distribution was deconvoluted to determine the relative contributions of the heavy flavor parent hadrons.
Fig.~\ref{VTX_pp} shows the ratio of electron yields from b-quarks,
and b + c quarks for p+p collisions~\cite{bib6}. The points with large
error bars are from $K\pi$ correlation measurements~\cite{bib14}.
Our new result is consistent with both the $K\pi$ correlation analysis
and with FONLL calculations. We also analyzed electrons
in Au+Au collisions to obtain the $R_{AA}$ for electrons from charm
and bottom quarks~\cite{bib6}.
The analysis assumed that the parent hadrons (e.g. D, B) follow the
$p_T$ distributions given by PYTHIA. The results demonstrate that the
Au+Au data are inconsistent with these input assumptions unless there
is also a large suppression of electrons from bottom across the measured
$p^e_T$ range. However,
this large suppression implies a change in the parent hadron $p_T$
distributions, which results in changes in the electron DCA distributions.
We are actively working on evaluation of the uncertainties caused by this
fact.

\section{Summary}
PHENIX has reported new findings in collisions at various energies and
species available at RHIC.
There is a significant enhancement of high $p_T$ hadrons and jets
in the peripheral d+Au collisions. The yield of $\psi\prime$ is
more heavily suppressed than that of $J/\psi$ in d+Au collisions.
A strong radial flow is seen in tip-tip enriched 0-2\,\% U+U collisions.
A sizable positive $v_1$ is observed in Cu+Au collisions.
In the same system, the $J/\psi$ yield measured in 1.2$<|\eta|<$2.2
as a function of $N_{part}$ shows larger suppression in the Cu-going side
than the Au-going side.
The space-time evolution of the ellipticity and triangularity of the Au+Au
collision system is observed in angular dependent HBT measurement.
PHENIX has seen an angular broadening of the away-side soft particles in
$\gamma$-h measurement, similar to h-h correlations and jet shape
measurements from RHIC and LHC.
The high $p_T$ hadron spectra showed that the fractional momentum shift under
the presence of hot dense matter is $\sim$25\,\% higher at LHC compared
to the RHIC top energy. Identification of electrons from bottom 
and charm quarks has been successful at RHIC using the VTX detector.
The ratio of ($b\rightarrow e$)/($b,c \rightarrow e$) for p+p collisions
is consistent with the previous measurement and well described by a FONLL
calculation.


\end{document}